\documentclass[a4paper,12pt]{article}
\usepackage{epstopdf}
\usepackage[utf8]{inputenc}
\usepackage[T1]{fontenc}
\usepackage{amsmath,amssymb}
\usepackage{graphicx}

\begin{document}
\textwidth=135mm
 \textheight=200mm
\begin{center}
{\bfseries Mott-Anderson freeze-out and the strange matter "horn"
}
\vskip 5mm
M. Naskr\c{e}t$^{\ast,\dag}$\footnote{Email: mnasket@cern.ch},
D. Blaschke$^{\dag,\ddag}$ and
A. Dubinin$^{\dag}$
\vskip 5mm
{\small {\it $^\ast$ CERN, 1211 Geneva 23, Switzerland}}\\
{\small 
{\it $^\dag$ IFT,
Uniwersytet Wroc\l{}awski, 50-204 Wroc\l{}aw, Poland}}\\
{\small 
{\it $^\ddag$ BLTP, Joint Institute for Nuclear Research, 141980 Dubna, Russia}} \\
\end{center}
\vskip 5mm
\centerline{\bf Abstract}
We discuss the $\sqrt{s}$-dependence of the $K^+/\pi^+$ ratio in heavy-ion collisions 
(the "horn" effect) within a Mott-Anderson localization model for chemical freeze-out.
The different response of pion and kaon radii to the hot and dense hadronic medium 
results in different freeze-out conditions.
We demonstrate within a simple model that this circumstance enhances the "horn" effect 
relative to statistical models with universal chemical freeze-out. 
\vskip 10mm

\section{\label{sec:intro}Mott-Anderson freeze-out for pions and kaons}

We like to investigate whether the Mott-Anderson model scenario 
\cite{Blaschke:2011ry,Blaschke:2011hm} 
would predict different chemical freeze-out conditions for kaons as compared to pions. 
According to this model the chemical freeze-out of hadrons from an expanding and cooling fireball
created in the course of an ultra-relativistic heavy-ion collision is based on the strong medium 
dependence of hadronic radii which govern hadron-hadron cross sections via
the geometrical Povh-H\"ufner law 
$\sigma_{hh'}=\lambda\langle r_h^2\rangle \langle r_{h'}^2\rangle$
 \cite{Povh:1987ju}.
The (inverse) collision time $\tau_{{\rm coll},h}^{-1}=\sum_{h'} \sigma_{hh'} n_{h'}$ determines the relaxation of 
hadron species $h$ towards its chemical equilibrium by reactive collisions with hadrons $h'$ having a 
number density $n_{h'}$. 
Due to the reduction of the chiral condensate in hot and dense hadronic matter the hadron radii swell 
as a precursor effect for the Mott-Anderson delocalization of hadron wave functions in the course of the
chiral restoration transition.
This behaviour has been quantified for the pion within the NJL model \cite{Hippe:1995hu} as
\begin{equation}
\label{rpi}
\langle r_\pi^2\rangle_{T,\mu} \simeq \frac{3}{4\pi^2} f_\pi^{-2}(T,\mu)
= \frac{3 M_\pi^2}{4\pi^2 m_q} \big| \langle \bar{q}q \rangle_{T,\mu}\big|^{-1}~. 
\end{equation}
In a hadronizing quark-gluon plasma the medium dependence of the chiral condensate has the form
\cite{Blaschke:2011ry,Blaschke:2011hm,Jankowski:2012ms}
\begin{equation}
\label{qqbar}
\langle\bar{q}q\rangle_{T,\mu} = \langle\bar{q}q\rangle^{MF}_{T,\mu}
+\sum_{h=M,B} \frac{\sigma_q^h}{m_q} n_h(T,\mu)~,
\end{equation}
where the first term stands for the quark meanfield contribution to the chiral condensate and the second one comes 
from the correlated quarks in hadrons with the scalar density
\begin{equation}
\label{ns}
n_h(T,\mu)=\frac{d_h}{2\pi^2}\int_0^\infty dk k^2 \frac{m_h}{E_h}\frac{1}{{\rm e}^{(E_h-\mu_h)/T}\mp 1}~.
\end{equation}
Here $h=M,B$ stands for (M)esons and (B)aryons with the minus (plus) sign for mesons (baryons) in  
Eq.~(\ref{ns}). 
The hadron sigma terms are defined as $\sigma_f^h=m_f(\partial m_h/\partial m_f)$ for the quark flavors
$f=u, d, s, \dots $ \cite{Jankowski:2012ms}.
The general expression (\ref{qqbar}) can also be applied for the strange quark condensate by replacing 
$q\leftrightarrow s$.

As shown in \cite{Blaschke:2011ry}, already a schematic resonance gas model consisting of $d_\pi=8$ pionic and $d_N=20$ 
nucleonic degrees of freedom gives an excellent description of the universal freeze-out line in the $T-\mu_B$ plane.
This curve has been given the parametric form \cite{Cleymans:2005xv} 
\begin{eqnarray}
\label{Tfo}
 T(\mu_B)&=&a-b\mu_B^2-c\mu_B^4\\
 \mu_B(\sqrt{s})&=&\frac{d}{1+e\sqrt{s}}~,
\label{mufo}
\end{eqnarray}
where $\sqrt{s}$ is the collision energy in the nucleon-nucleon center of mass system. 
We will identify this curve with the Mott-Anderson freeze-out line for pions. 

In order to apply the model to kaon freeze-out we have to consider the in-medium kaon radius which 
in analogy to (\ref{rpi}) reads
\begin{equation}
\label{rK}
\langle r_K^2\rangle_{T,\mu} \simeq \frac{3}{4\pi^2} f_K^{-2}(T,\mu)
= \frac{3 M_K^2}{\pi^2 (m_q+m_s)} \big|  \langle \bar{q}q \rangle_{T,\mu}+\langle \bar{s}s \rangle_{T,\mu}\big|^{-1}~. 
\end{equation}
Now we are in the position to understand why the Mott-Anderson model predicts different freeze-out lines for pions and kaons. 
While the pion radius directly responds to the medium dependence of the light quark condensate which gets modified already 
due to the presence of pions themselves as the lightest species in the system, the kaon radius depends on the strange quark 
condensate (see Eq.~(\ref{qqbar}) for $q\leftrightarrow s$) which is more inert to the medium since its modification requires 
that strangeness-carrying species be abundant. 
The latter, however, are suppressed relative to light hadrons by their larger mass so that a possible approximation reads
\begin{equation}
\label{ssbar}
\langle\bar{s}s\rangle_{T,\mu} \approx \langle\bar{s}s\rangle^{MF}_{T,\mu}~.
\end{equation}
The numerical results for the pion and kaon freeze-out lines are shown in the right panel of the final 
Fig.~\ref{fig:freezeout} below. 
 
\section{The kaon freeze-out line from $K^+/\pi^+$ data} 
\subsection{Pion and strangeness chemical potentials}
To address the strange matter "horn'' we have to consider the ratio of yields
of positively charged kaons to positively charged pions, denoted as 
$K^+/\pi^+=n_{K^+}/n_{\pi^+}$.
Within the thermal statistical model, the number densities of charged pions 
and kaons are given by
\begin{eqnarray}
\label{statistics}
 n_{\pi^+}&=& n_{\pi^-} = \int_0^\infty \frac{dp}{2\pi^2}p^2
\frac{1}{{\rm e}^{(\sqrt{p^2+m_{\pi}^2}-\mu_\pi)/{T}}-1}~,
\\
 n_{K^\pm}&=&\int_0^\infty \frac{dp}{2\pi^2}p^2
\frac{1}{{\rm e}^{(\sqrt{p^2+m_{K}^2}\mp(\mu-\mu_s))/{T}}-1}
~,
\label{yields}
\end{eqnarray}
where $\mu=\mu_B/3$ is the light quark chemical potential.
For the strange quark chemical potential $\mu_s$  which has to assure vanishing 
net strangeness we take $\mu_s=\varepsilon \mu$, being guided by \cite{Karsch:2010ck} 
to adopt a proportionality to $\mu$. 

\begin{figure}[!htb]
\includegraphics[width=0.5\textwidth]{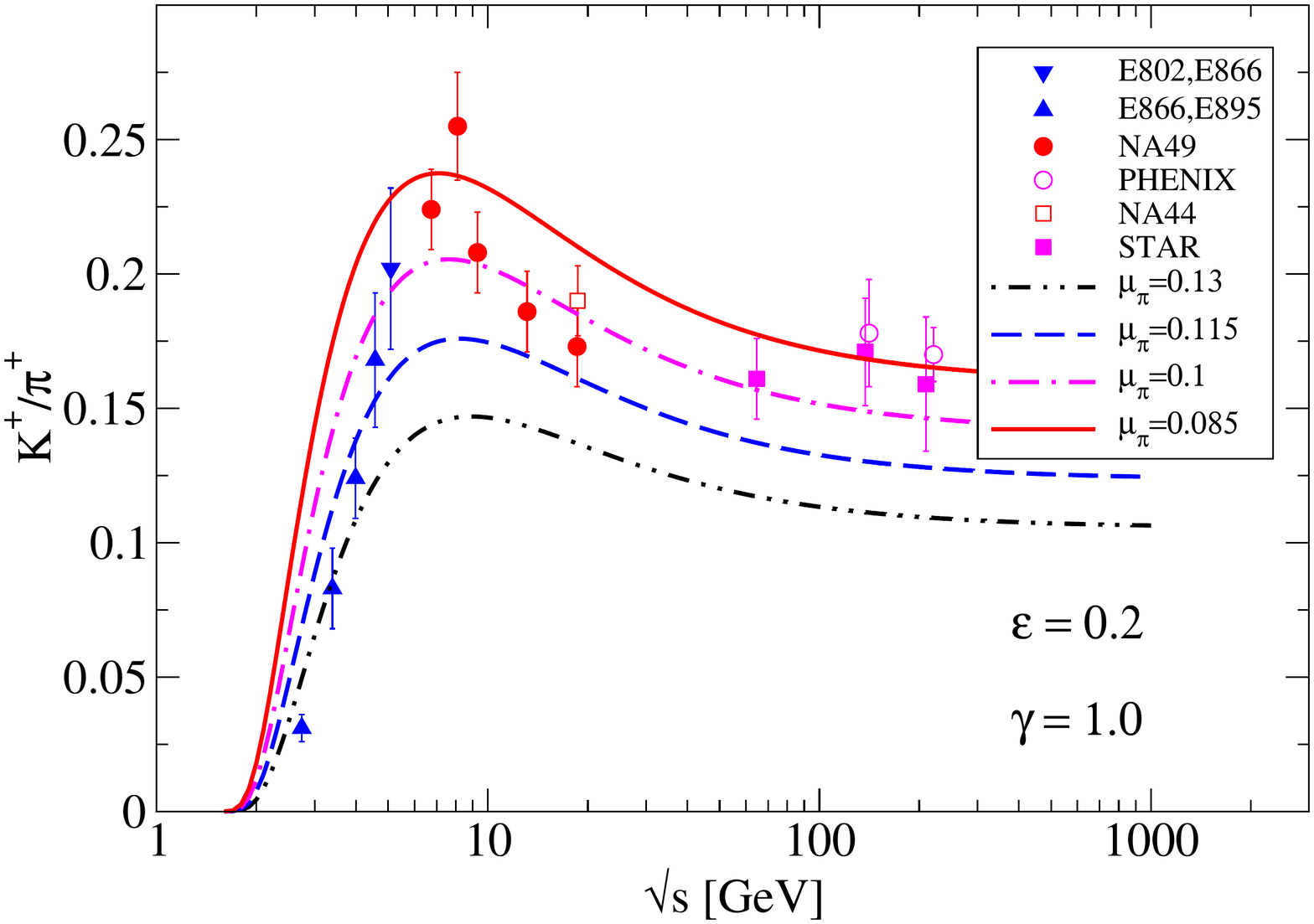}
\includegraphics[width=0.5\textwidth]{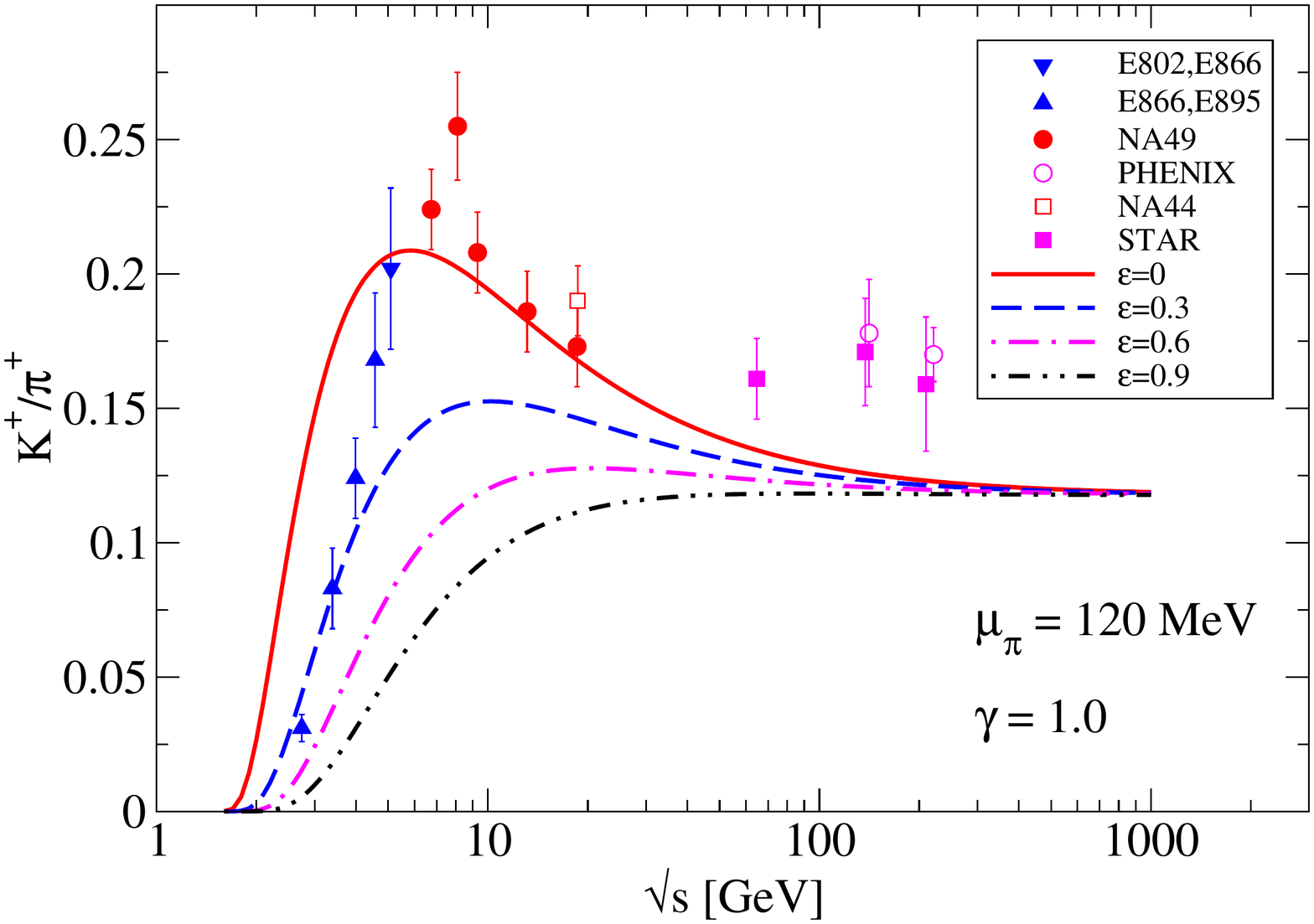}
\caption{The ratio $K^+/\pi^+=n_{K^+}/n_{\pi^+}$ according to Eqs. 
(\ref{statistics})-(\ref{yields}) compared with experimental data 
for varying nonequilibrium parameter $\mu_\pi$ (left) and varying strangeness
chemical potential $\varepsilon$ (right). 
\label{fig:kpi}}
\end{figure}

Forming the ratio $K^+/\pi^+=n_{K^+}/n_{\pi^+}$ one can now attempt a preliminary 
comparison with data and explore the role of the two parameters $\mu_\pi$ and $\varepsilon$,
see Figs.~\ref{fig:kpi}.

The chemical potential $\mu_\pi$ for pions was introduced 
as a phenomenological parameter to decribe the low-momentum enhancement of the
pion distribution observed at CERN SPS \cite{Kataja:1990tp}, with a value of 
the order of the pion mass signalling the onset of pion condensation.
Recently, the parameter $\mu_\pi$  has been rediscovered as a possibility to solve the 
proton-to-pion puzzle at LHC \cite{Begun:2013nga}.
Here it has been given the meaning of  parameter characterizing the nonequilibrium 
nature of the pion distribution as imprinted in their transverse momentum spectra.
Note that role of a pion chemical potential in the context of condensation phenomena
has been pointed out, e.g., in Ref~\cite{Turko:1993dy}.
An elucidation of the emergence of this parameter from an underlying nonequilibrium 
description of pion production in heavy-ion collisions is, to best of our knowledge, still
missing.
\begin{figure}[!htb]
\includegraphics[width=0.5\textwidth]{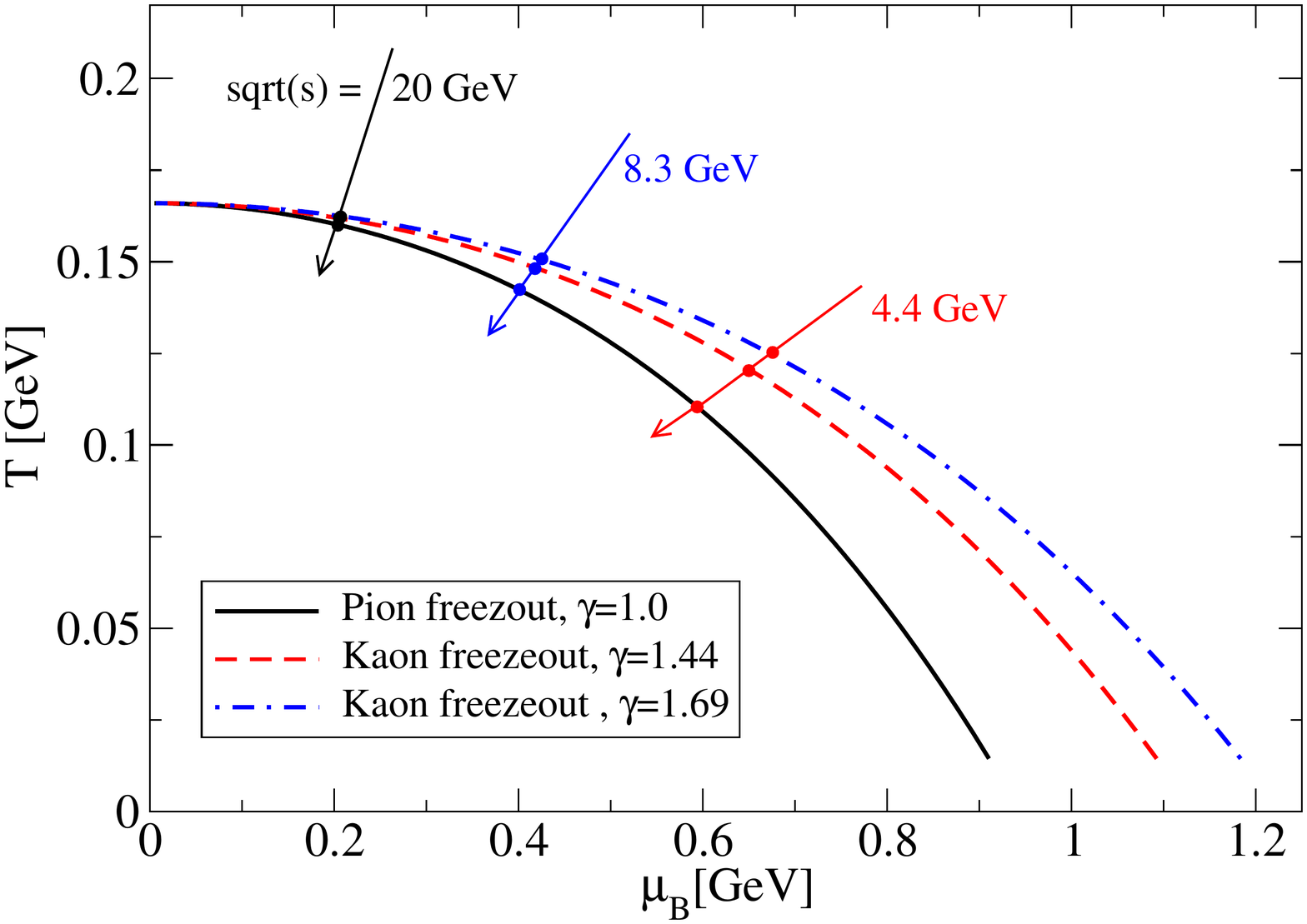}
\includegraphics[width=0.5\textwidth]{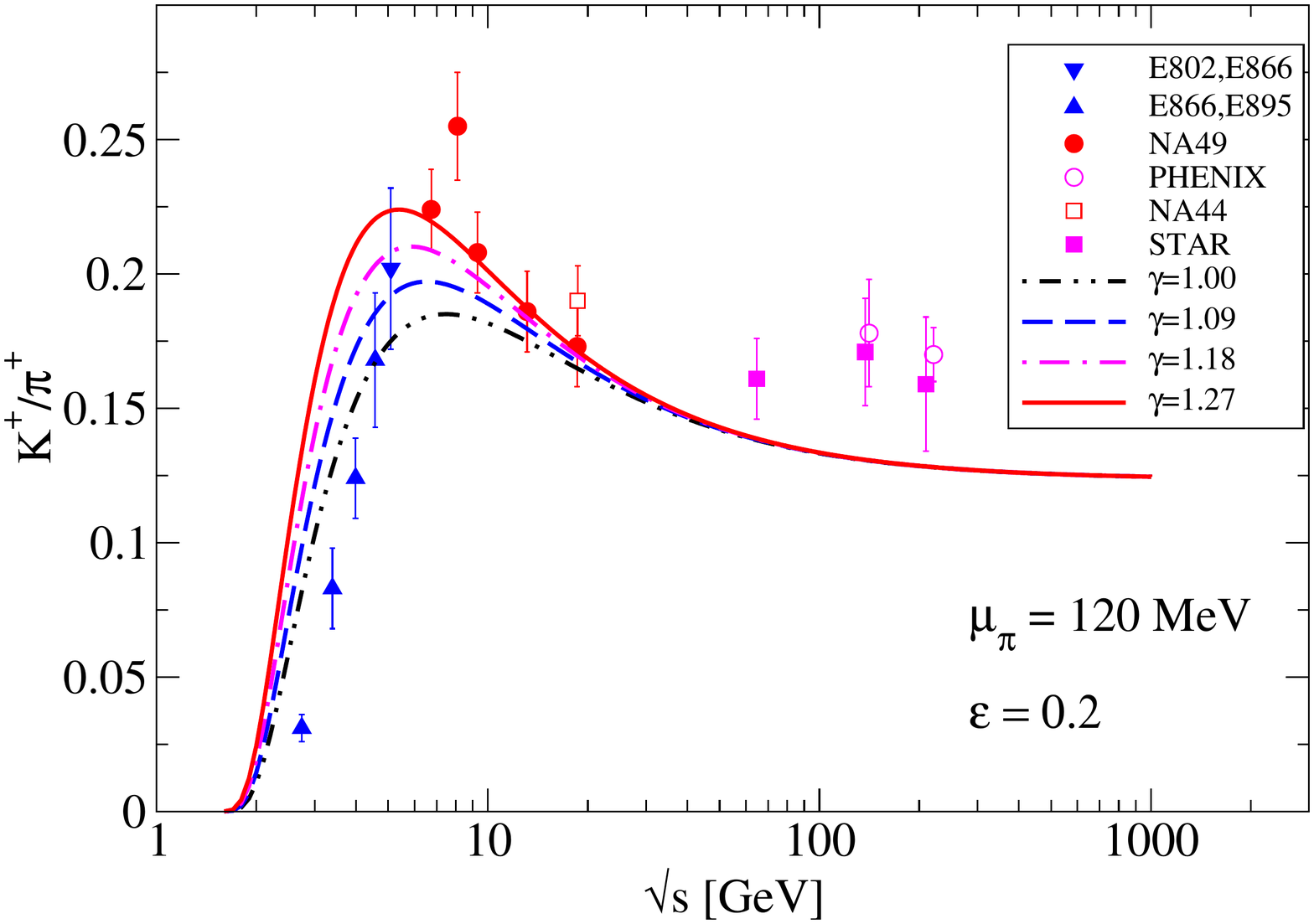}
\caption{Left panel: The difference between the freeze-out lines for kaons $T_{fo}^K(\mu)$
and for pions $T_{fo}^\pi(\mu)$ (solid line) is being quantified by the parameter $\gamma$. 
Two lines with $\gamma>1.0$ are shown for comparison. 
Right panel: $K^+/\pi^+$ ratio as a function of $\sqrt{s}$ for different values of $\gamma$ compared
to the data (symbols).
\label{fig:gamma}
}
\end{figure}

\subsection{Parametrizing different kaon and pion freeze-out}
In a next step, we want to explore the possibility that the kaon freeze-out 
points in the $T-\mu_B$ plane do not coincide with those for the pions which 
are assumed to lie on the universal freeze-out curve parametrized by 
Eqs.~(\ref{Tfo}) and (\ref{mufo}).
To this end, we make the ansatz
\begin{eqnarray}
\mu_{fo}^{K}(T) = \sqrt{\gamma} ~\mu_{fo}^{\pi}(T)~, 
\end{eqnarray}
where $\mu_{fo}^{\pi}(T)$ is a solution of the biquadratic equation (\ref{Tfo})
\begin{eqnarray}
\mu_{fo}^{\pi}(T)=\sqrt{\frac{b}{2c}\left(\sqrt{1+\frac{4c}{b^2}(T_0-T)}-1 \right)} ~.
\end{eqnarray}

We have denoted $T_0=a$ in order to make clear the meaning of the parameter 
$a$ as the freeze-out temperature at vanishing baryochemical potential.
The parameter $\gamma\ge 1$ reflects our expectation from the Mott-Anderson model of 
Sect.~\ref{sec:intro} that the freeze-out line for the kaons lies for a given temperature 
at larger $\mu_B$-values than that for pions, see Fig.~\ref{fig:gamma}, left panel.

The assumption we make now is that the evolution of the system across the 
freeze-out lines follows lines of constant entropy $T/\mu=\kappa$ for a given 
$\sqrt{s}$, and 
\begin{equation}
\label{kappa}
\kappa(\sqrt{s})=T(\sqrt{s})/\mu(\sqrt{s})
\end{equation} 
can be found from the ratio of Eqs.~(\ref{Tfo}) and (\ref{mufo}).
In order to determine for given $\sqrt{s}$ the freeze-out point for kaons, one
finds first $\mu_{fo}^K$ from solving the fourth order equation in $\mu$
\begin{eqnarray}
\mu^4 +\frac{b\gamma}{c}\mu^2+ \frac{\gamma^2~\kappa(\sqrt{s})}{c} \mu 
-\frac{\gamma^2 T_0}{c}=0~,
\end{eqnarray}
and then inserts the solution in (\ref{kappa}) to find the 
kaon freeze-out temperature
\begin{equation}
\label{TfoK}
T_{fo}^K(\sqrt{s})=\kappa(\sqrt{s})~\mu_{fo}^K(\sqrt{s})~.
\end{equation} 

\begin{figure}[!ht]
\includegraphics[width=0.5\textwidth]{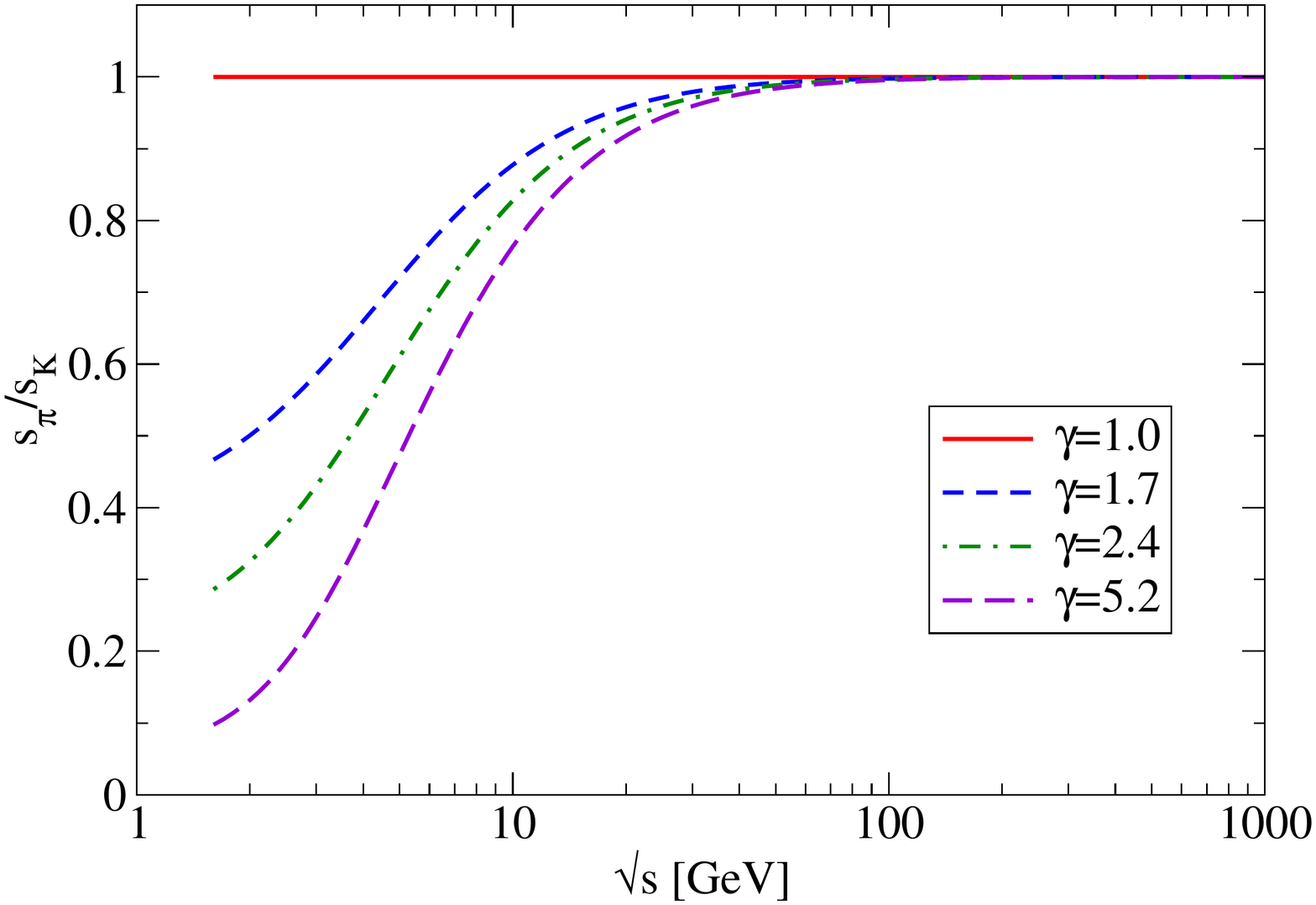}
\includegraphics[width=0.5\textwidth]{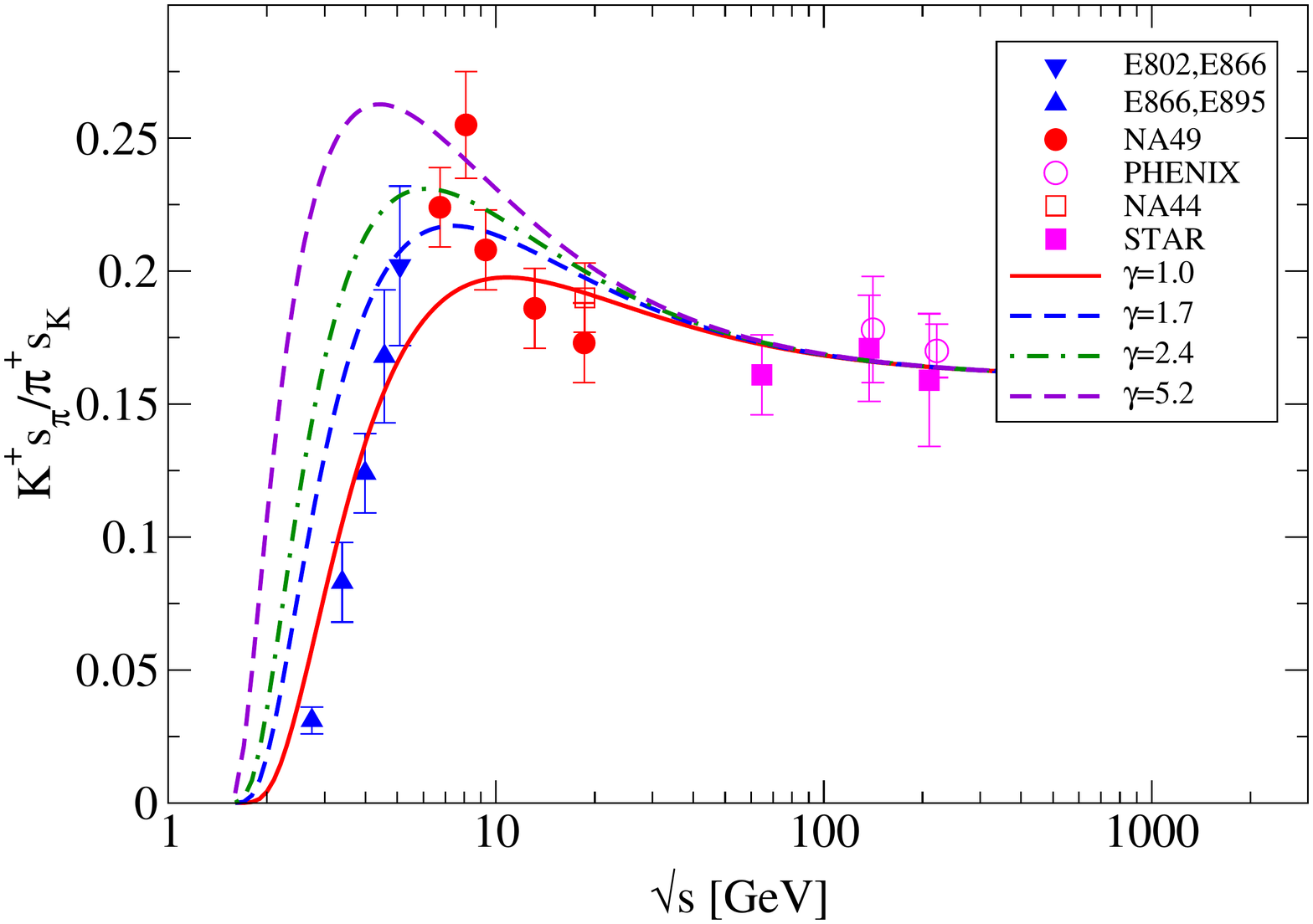}
\caption{Left panel: ratio of entropy densities 
as a measure for the change in the freeze-out volume between kaon and pion freeze-out. 
Right panel: $K^+/\pi^+$ ratio for different values of $\gamma$ including the volume 
expansion effect.
\label{fig:entropy}
}
\end{figure}

As a result of this procedure we obtain the ratio $K^+/\pi^+$ shown in Fig.~\ref{fig:entropy},
where kaon and pion densities are to be taken at their respective, differing 
freeze-out points which are related by a straight line in the $T-\mu_B$ plane. 
Varying $\sqrt{\gamma}$ in the range $1\le \sqrt{\gamma} \le 1.3$ we observe 
that the slope of the $K^+/\pi^+$ ratio below the ``horn'' gets diminished 
while the asymptotics at high $\sqrt{s}$ remains unaffected. 

\begin{figure}[!htb]
\includegraphics[width=0.5\textwidth]{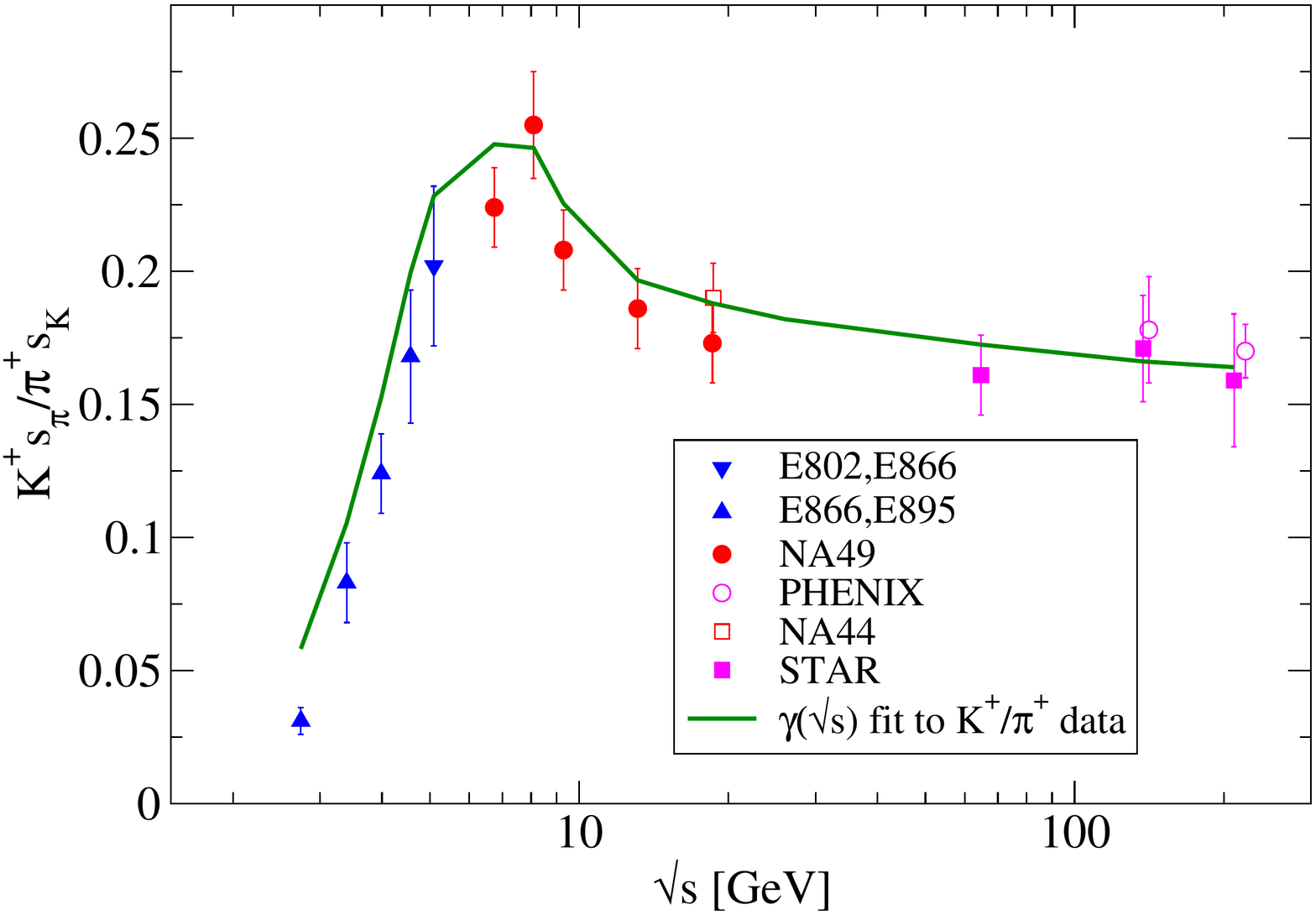}
\includegraphics[width=0.5\textwidth]{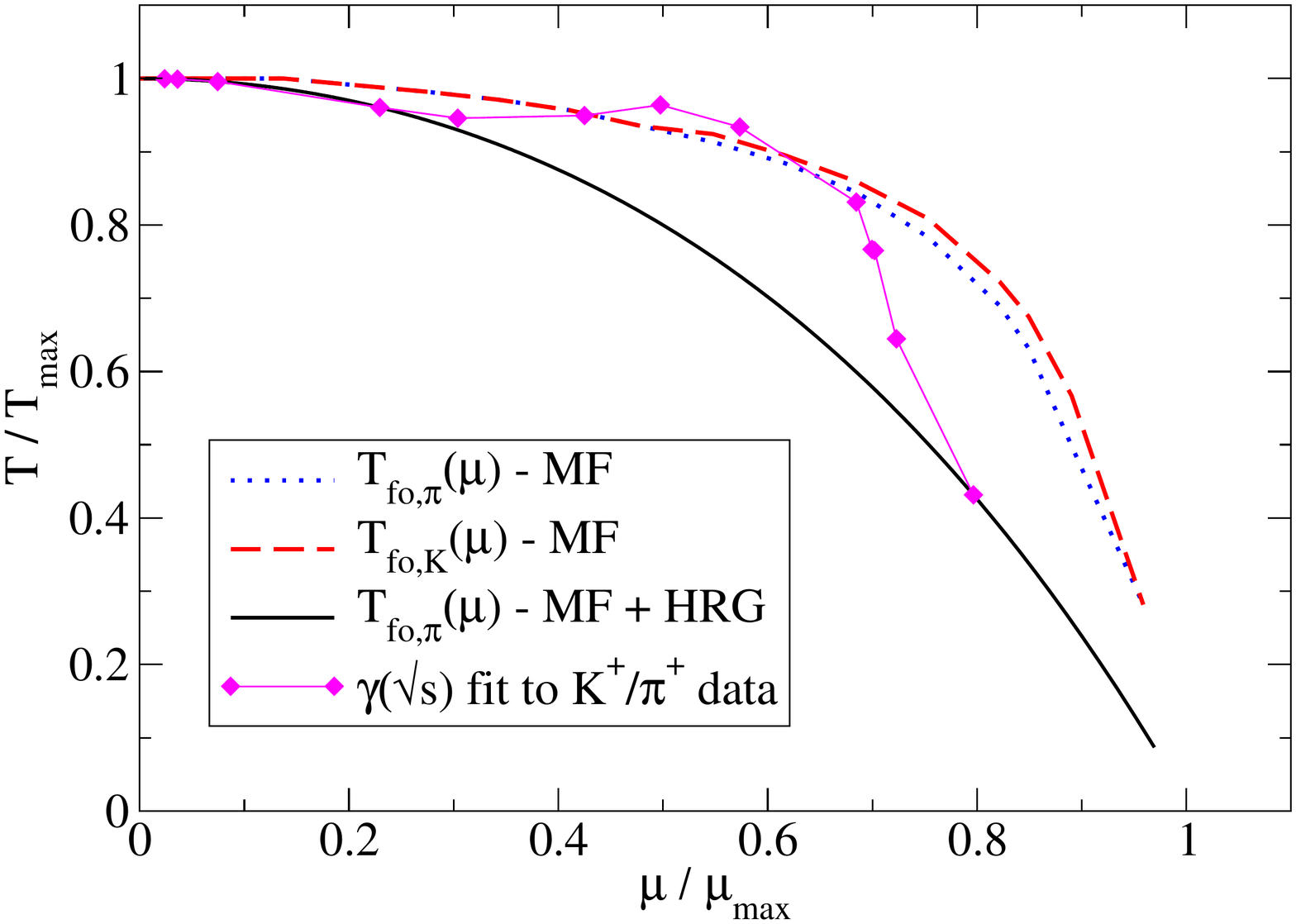}
\caption{Left panel: 
Fit of $\gamma(\sqrt{s})$ (line) to the experimental $K^+/\pi^+$ ratio (symbols). 
Right panel: Freeze-out lines of kaons and pions in the $T-\mu$ plane from the 
Mott-Anderson localization model compared with values extracted from the 
experimental $K^+/\pi^+$ ratio by parametrizing the $\gamma(\sqrt{s})$ of the toy 
model.
\label{fig:freezeout}}
\end{figure}

\section{Result and discussion}
In Fig.~\ref{fig:freezeout} we show in the left panel the result for the $K^+/\pi^+$ ratio
with a pointwise fit of the function  $\gamma(\sqrt{s})$ in order to describe the shape 
of the "horn" effect. This parametrizes the deviation of the kaon freeze-out line from that 
of the pions in the $T-\mu$ plane as shown in the right panel of that figure.
For not too high values of $\mu<0.7\ \mu_{\rm max}$ the comparison with the prediction from the 
Mott-Anderson model works very well. Beyond this value of the chemical potential, i.e., 
for energies below the peak position of the "horn" at $\sqrt{s}\sim 8$ GeV, both the simple assumption 
of neglecting strange hadrons in the formula for the strange condensate in the Mott-Anderson model 
and the simplifications of the toy model for extracting the kaon freeze-out line break down.
Among the issues to be included upon improvement are resonance decays, excluded volume effects,
the canonical statistical suppression factor and an elucidation of the interrelation between 
the non-universal freeze-out discussed here and the strangeness enhancement factors used in 
alternative approaches \cite{Becattini:2012sq,Petran:2013qla,BraunMunzinger:2001ip}.
Moreover, it will be interesting to study the relationship to other approaches to the "horn"
effect like, e.g., the modified statistical model \cite{Sagun:2014sya} or the "quarkyonic" explanation of 
Ref.~\cite{Andronic:2009gj}.   
Nevertheless, the first step gave a promising result and the necessary improvements to 
go further can clearly be identified.

\subsection*{Acknowledgements}
We are grateful to Krzysztof Redlich and Ludwik Turko for their critical remarks concerning
this work. We thank Jakub Jankowski for discussions in an early stage of this work.
M.N. acknowledges support by NCN under grant number UMO-2012/04/M/ST2/00816 and 
by NA61/SHINE for his participation in experiments at CERN Geneva. 
D.B. thanks all participants of the ECT* Trento workshop on "QCD Hadronization and 
the Statistical Model" for inspiring discussions. His work was supported by NCN under 
grant number UMO-2011/02/A/ST2/00306.  
A.D. received support by NCN under grant number UMO-2013/09/B/ST2/01560.

\end{document}